\documentclass[prd,twocolumn,showpacs,amsmath,amssymb]{revtex4}

\usepackage{amssymb}
\usepackage{mathrsfs}
\usepackage{txfonts}

\usepackage{graphicx}
\usepackage{dcolumn}% Align table columns on decimal point
\usepackage{bm}% bold math

\begin{document}

\title{Primordial non-Gaussianity in warm inflation using $\mathbf{\delta N}$ formalism}

\author{Xiao-Min Zhang}
%\thanks{Corresponding author}
\email{zhangxm@mail.bnu.edu.cn}
\affiliation{School of Science, Qingdao University of Technology, Qingdao 266033, China}
\author{Hong-Yang Ma}
\affiliation{School of Science, Qingdao University of Technology, Qingdao 266033, China}
\author{Peng-Cheng Chu}
\affiliation{School of Science, Qingdao University of Technology, Qingdao 266033, China}
\author{Jun-Ting Liu}
\affiliation{Qindao College, Qingdao University of Technology, Qingdao 266033, China}
\author{Jian-Yang Zhu}
\thanks{Corresponding author}
\email{zhujy@bnu.edu.cn}
\affiliation{Department of Physics, Beijing Normal University, Beijing 100875, China}

\date{\today}
%------------------------------------------------------------------------------------------------------------
\begin{abstract}
 A $\delta N$ formalism is used to study the non-Gaussianity of the primordial curvature perturbation on an uniform density hypersurfaces generated by the warm inflation for the first time. After introducing the framework of the warm inflation and the $\delta N$ formalism, we obtain an analytic expression for the nonlinear parameter $f_{NL}$ that describes the non-Gaussianity in slow roll approximation, and find that the $\delta N$ formalism gives a very good result. We analyse the magnitude of $f_{NL}$ and compare our result with those of the standard inflation. Then we discuss two concrete examples: the quartic chaotic model and the hilltop model. The quartic potential model can again be in very good agreement with the Planck results in the warm inflationary scenario, and we give out the concrete results of how the nonlinear parameter depends on the dissipation strength of the warm inflation and the amounts of expansion. We find that the range of the nonlinear parameters in these two cases are both well inside of the allowed region of Planck.
\end{abstract}
\maketitle

%------------------------------------------------------------------------------------------------------------
\section{\label{sec1}Introduction}
Inflation is an important branch of cosmology which can successfully solve the problems such as the horizon, the flatness and the monopole \cite{Guth1981,Linde1982,Albrecht1982}. What's more, the inflation can provide a graceful mechanism to clarify the observed anisotropy of the cosmological microwave background (CMB) and the large scale structure exactly \cite{Weinberg,LiddleLyth,Dodelson,Bassett2006}. Till now there are two kinds of inflationary theory - the standard inflation and the warm inflation. The warm inflation was first proposed by Berera etc in 1995 \cite{BereraFang,Lisa2004,Berera2000}, and then we often call the standard inflation that was first proposed in 1981 \cite{Guth1981,Linde1982,Albrecht1982} as the cold inflation. The warm inflation inherits the advantages of the standard inflation such as solving horizon, flatness and monopole problems and generate nearly scale-invariant power spectrum, but also has some improvements and distinctions. A most important distinction is the origin of the density fluctuations. The perturbations naturally arise from the vacuum quantum fluctuations in standard inflation \cite{Weinberg,LiddleLyth,Dodelson,Bassett2006} or the thermal fluctuations in warm inflation \cite{BereraFang,Lisa2004,Berera2000}. The warm inflation can cure the ``$\eta$'' problem \cite{etaproblem} and the problem of overlarge amplitude of the inflaton suffered in some standard inflationary models \cite{Berera2005,BereraIanRamos}. In addition, the slow roll approximation can much easily to be satisfied in warm inflation thanks to the thermal damped effect \cite{Ian2008,Campo2010,ZhangZhu,Zhang2014}. The warm inflation has developed significantly in the past twenty years and broad the scope of inflationary theory. In recent years, many works concentrate on the first principle realization of the warm inflation \cite{Berera1999,MossXiong2006}, the thermal fluctuations of warm inflation \cite{Lisa2004,Berera2000,Taylor2000,Chris2009,MossXiong}, the dissipation coefficient in different cases \cite{Zhangyi2009,MarGil2013,Mar2007,Ramon2013,Ramon2014}, and the consistency issue of warm inflation \cite{Ian2008,Campo2010,ZhangZhu,Zhang2014,ZhangTachyon}. What's more, some models that were excluded by Planck observations \cite{PLANCKI2015} in standard inflation can again be in very good agreement with the Planck results in warm inflationary scenario. One representative example is the quartic chaotic potential model \cite{Sam2014}, which we'll discuss in detail hereinafter.

In the study of inflation, one typically calculates the power spectrum of scalar perturbations and the amplitude of gravitational waves. This information is very important, however, it contains only two-point correlation statistics information and therefore is too limited to discriminate among a large range of inflationary models. This is the degeneracy problem (i.e., a single set of observables maps to a range of different inflation models)\cite{Eassona2013} in inflation theory. Even a precise measurement of the spectral index, the running of spectral index, and the detection of gravitational wave will not allow us efficiently discriminate among them. As a result, there has been growing interest in calculating higher-order correlation functions. If a perturbation quantity is exact Gaussian distributed, the two-point correlation function or its Fourier transform, the power spectrum is all that is needed in order to completely characterize it from a statistical point of view \cite{Bartolo2004}. So the three-point function, or its Fourier transform, the Bispectrum represents the lowest order statistics able to distinguish non-Gaussian from Gaussian perturbations \cite{Heavens1998,Ferreira1998}. In this paper we only concentrate on the lowest order non-Gaussianity.

The primordial curvature perturbation in slow-roll inflationary scenario is dominated by the Gaussian term. Many works has been concentrated on the issue of the non-Gaussianity in standard inflation in both canonical inflationary theory \cite{Gangui1994,Calzetta1995,David2005,Bartolo2004,Lyth2005,Zaballa2005} and noncanonical theory \cite{Creminelli2003,Tong2004}. The single canonical slow-roll inflationary model itself produces negligible non-Gaussianity, but the non-Gaussianity can be significant in noncanonical or multi-field inflationary models. In recent years, the non-Gaussianity generated in both strong and weak warm inflationary models was performed in many works such as Refs.\cite{MossXiong,Gupta2006,Gupta2002,MarGil2014}. The non-Gaussianity in noncanonical warm inflation was considered in \cite{Zhang2015} and find that small sound speed and large dissipation strength can both enhance the magnitude of the non-Gaussianity. About ten years ago, $\delta N$ formalism was used to calculate the issue of non-Gaussianity for the first time \cite{David2005,Zaballa2005}. This formalism is straight and convenient to use in calculating the non-Gaussianity and was often used in multi-field cold inflationary theory \cite{Vernizzi2006,Battefeld2007,Tower2010}. In this paper we try to take this formalism to calculate the primordial non-Gaussianity generated in warm inflation. The calculations about non-Gaussianity in kinds of models are almost in the context of the first-order cosmological perturbation theory, and the analysis of the second-order cosmological perturbation theory is really complicated. Calculations about the second-order perturbation theory were done in some works \cite{Pyne1996,Komatsu2001,David2005,Lyth2005} and yield a model-independent second-order result $f_{NL}\sim \mathcal{O} (1)$. We'll discuss later when the second order cosmological perturbation result will play important role. Also we'll check our results with the latest Planck result \cite{PLANCKNG2015}.

The paper is organized as follows: In Sec. \ref{sec2}, we introduce the warm inflationary scenario briefly and review the basic equations of the picture. In Sec. \ref{sec3}, we introduce the $\delta N$ formalism to calculate the nonlinear parameter $f_{NL}$ and give out the result for warm inflationary scenario. Then we discuss the non-Gaussian signature in quartic chaotic potential model and hilltop model respectively in Sec. \ref{sec4}. Finally, we draw the conclusions in Sec. \ref{sec5}.

\section{\label{sec2} warm inflation dynamics}

In warm inflation, the scalar inflaton field is not isolated but has interactions with other sub-dominated fields. Due to the interactions, a significant amount of radiation was produced constantly during the inflationary epoch, so the Universe is hot with a non-zero temperature $T$. The necessary conditions for this to happen were discussed in \cite{Chris2008}.
The total matter action of the multi-component Universe in warm inflation is
\begin{equation}\label{action}
  S=\int d^4x \sqrt{-g}  \left[ \mathcal{L}_{\phi}+\mathcal{L}_R+\mathcal{L}_{int}\right],
\end{equation}
where the Lagrangian density of the inflaton field is $\mathcal{L}_{\phi}$, $\mathcal{L}_R$ is the Lagrangian density of the radiation fields, and $\mathcal{L}_{int}$ denotes the interactions between the scalar fields. In the Friedmann-Robertson-Walker (FRW) Universe, the mean inflaton field is homogeneous, i.e. $\phi=\phi(t)$. By varying the action with respect to the inflaton field and making some assumptions and calculations \cite{BereraFang,Berera1999,Zhang2014}, we can get the evolution equation of the inflaton field:
\begin{equation}
\ddot{\phi}+(3H+\Gamma )\dot{\phi}+V_{eff,\phi} =0  \label{EOMphi},
\end{equation}
where $H$ is the Hubble parameter with the Friedmann equation:
\begin{equation}\label{Friedmann}
  3H^2=8\pi G\rho.
\end{equation}
In Eq. (\ref{EOMphi}), $\Gamma$ is the dissipation coefficient, $V_{eff,\phi}$ is the effective potential acquired thermal corrections and the subscript $\phi$ denotes a derivative. The effective potential $V_{eff,\phi}$ is different from the zero-temperature potential in cold inflation, but the thermal correction is small \cite{Ian2008,Campo2010,Zhang2014}. For simplicity we'll write $V_{eff}$ as $V$ hereinafter. The term $\Gamma \dot{\phi}$ describes the dissipation effect of $\phi $ to radiations \cite{BereraFang,Berera2005,Berera2000,BereraIanRamos}, which is a thermal
damping term. For simplicity, $\Gamma $ is often
set to be a constant in some papers \cite{Herrera2010,Xiao2011,Taylor2000}.
Considering some concrete models of the interaction between inflaton and
other fields, different form of $\Gamma $ can been obtained \cite{MossXiong2006,MarGil2013,BereraIanRamos} and was found more likely to be a function of the inflaton field and even the temperature.

An important parameter in warm inflationary theory is the dissipation strength, which describes the efficiency of warm inflation and is defined as
\begin{equation}\label{r}
  r=\frac{\Gamma}{3H}
\end{equation}
where $r\gg1$ corresponds to a strong regime of warm inflation and $r\ll1$ corresponds to a weak regime of warm inflation.

The thermal dissipative effect of warm inflation is associated with the production
of entropy. The expression for entropy density from thermodynamics is $s=-\partial f/\partial T$, and we have $s\simeq -V_T$ (the subscript $T$ denotes a derivative) for during inflation the free energy $f=\rho -Ts$ is dominated by
the potential.

The total energy density of the multi-component Universe is
\begin{equation}
\rho =\frac 12\dot{\phi}^2+V(\phi ,T)+Ts,  \label{rho}
\end{equation}
and the total pressure is given by
\begin{equation}
p=\frac 12\dot{\phi}^2-V(\phi ,T).  \label{p}
\end{equation}
The energy-momentum conservation
\begin{equation}
\dot{\rho}+3H(\rho +p)=0
\end{equation}
combining with Eq. (\ref{EOMphi}) yields the entropy production equation
\begin{equation}
T\dot{s}+3HTs=\Gamma \dot{\phi}^2.  \label{entropy}
\end{equation}
The equation above is equivalent to the radiation energy density producing
equation
\begin{equation}
\dot{\rho}_r+4H\rho _r=\Gamma \dot{\phi}^2,
\end{equation}
when the thermal correction to the effective potential is little enough (this can be ensured by the slow roll conditions).

The exact solution of these equations above are difficult to obtain, so during inflation we often apply a slow roll approximation to drop the highest derivative terms in the evolution equation and Friedmann equation,
\begin{equation}
\dot{\phi}=-\frac{V_\phi }{3H(1+r)},  \label{SRdotphi}
\end{equation}
\begin{equation}
Ts=r\dot{\phi}^2,  \label{SRTs}
\end{equation}
\begin{equation}
H^2=\frac{8\pi G}3 V,  \label{SRH}
\end{equation}
\begin{equation}
4H\rho _r=\Gamma \dot{\phi}^2.  \label{SRrho}
\end{equation}
The validity of the slow roll approximation depends on the slow roll conditions given by stability analysis \cite{Ian2008,Campo2010,Lisa2004,Zhang2014}. The slow roll conditions are associated with some important slow roll parameters defined as
\begin{equation}
\epsilon =\frac{M_p^2}{2}\left(\frac{V_{\phi}}{V}\right) ^2, \eta =M_p^2\frac {V_{\phi \phi}}{V}, \beta
=M_p^2\frac{V_{\phi}\Gamma_{\phi}}{V\Gamma},
\end{equation}
The slow roll approximation holds when $\epsilon\ll 1+r$, $\eta\ll1+r$ and $\beta\ll1+r$. Any quantity of order $\epsilon/(1+r)$ will be described as being first order in the slow roll approximation.
When the dissipation coefficient depends on temperature, there will exist two additional slow roll parameters: $b=\frac {TV_{\phi T}}{V_{\phi}}$ and $c=\frac{T\Gamma_T}{\Gamma}$. The slow roll validity requires: $b\ll1$, $|c|<4$ \cite{Ian2008,Campo2010}.

When the slow roll parameter $\epsilon \simeq 1+r$, $\ddot{a}=0$, which implies the end of the inflationary
phase. The number of e-folds in warm inflation is given by
\begin{equation}
N(\phi)=\int Hdt=-\frac 1{M_p^2}\int_\phi ^{\phi _e}\frac V{V_\phi} (1+r)d\phi,  \label{efold}
\end{equation}
where $M_p^2=\frac 1{8\pi G}$ and the subscript $e$ is used to denote the end of inflation.

\section{\label{sec3}Non-Gaussianity in warm inflation}

\subsection{\label{sec31}Introduction of $\delta N$ formalism}
Now we introduce a general procedure for calculating the level of non-Gaussianity by means of $\delta N$ formalism. This formalism was proposed in \cite{Lyth2005,LythMalik,Starobinsky,Sasaki1996,Sasaki1998} and then was often used in calculating the non-Gaussianity of double and multi-field inflationary model \cite{Vernizzi2006,Battefeld2007,Tower2010}.

The primordial curvature perturbation on an uniform density hypersurfaces of the Universe, denoted by $\zeta$, is already present a few Hubble times before cosmological scales start to enter the horizon. And the perturbation $\zeta$ was found to be Gaussian term dominated with a nearly scale-invariant spectrum.

Considering little perturbations, the Universe can be described as a quasi-homogeneous flat FRW spacetime with scale factor $a(t)$. The spatial metric is given by \begin{equation}\label{gij}
  g_{ij}=a^2(t)e^{2\zeta(t,\mathbf{x})}\gamma_{ij}(t,\mathbf{x})=\tilde{a}^2(t,\mathbf{x})\gamma_{ij}(t,\mathbf{x}),
\end{equation}
where $\gamma_{ij}(t,\mathbf{x})$ has unit determinant and accounts for the tensor perturbation. We can find that according to this definition, $\zeta$ is the perturbation in $\ln a$.

We consider a slicing whose metric has the form in Eq. (\ref{gij}) without the $\zeta$ factor, which we call the flat slicing. Starting from any initial flat slice at time $t_{in}$, one can define the amount of expansion $N(t,\mathbf{x})\equiv\ln\left[\frac{\tilde{a}(t)}{a(t_{in})}\right]$ to a final slice of uniform energy density and unperturbed amount of expansion $N_0(t)\equiv\ln\left[\frac{a(t)}{a(t_{in})}\right]$. Then according to $\delta N$ formalism \cite{Starobinsky,Sasaki1996,Sasaki1998}, we have
\begin{equation}\label{zeta}
  \zeta(t,\mathbf{x})=\delta N \equiv N(t,\mathbf{x})-N_0(t) .
\end{equation}

During inflation, the evolution of the Universe is supposed to be determined by one or more scalar fields. For simplicity, we choose the flat slicing gauge, and considering perturbations, each scalar field can be expanded as $\phi_i(t,\mathbf{x})=\phi_i(t)+\delta\phi_i(t,\mathbf{x})$. As we mentioned above, the curvature perturbation $\zeta$ is nearly Gaussian, so we can have a good accuracy expression for $\zeta$ up to second order
\begin{equation}\label{zeta2}
  \zeta(t,\mathbf{x})=\delta N\simeq \sum_i N_{,i}(t)\delta\phi_i+\frac12\sum_{ij} N_{,ij}\delta\phi_i\delta\phi_j,
\end{equation}
where $N_{,i}\equiv\frac{\partial N}{\partial\phi_i}$ and $N_{,ij}\equiv\frac{\partial^2 N}{\partial\phi_i\partial\phi_j}$. They may be entirely responsible for any observed non-Gaussianity if the scalar field perturbations are Gaussian to sufficient accuracy. Fortunately, the inflaton field perturbation in warm inflation is almost Gaussian, which we will see below. In warm inflation, we expand the full inflaton as $\Phi(\mathbf{x},t)=\phi(t)+\delta\phi(\mathbf{x},t)$, with $\delta\phi$ is the small perturbation around the homogenous background field $\phi(t)$. The evolution equation of the leading order inflaton perturbation (denoted as $\delta\phi_1$) in warm inflation is \cite{Berera2000,Zhang2014}
\begin{eqnarray}\label{deltaphi1}
\frac{d}{dt}\delta\phi_1(\mathbf{k},t)&=&\frac{1}{3H+\Gamma}\left[-k^2\delta\phi_1(\mathbf{k},t)\right.\nonumber\\
&-&\left.V_{\phi\phi}(\phi(t))\delta\phi_1(\mathbf{k},t)+\xi(\mathbf{k},t)\right],
\end{eqnarray}
where $\xi$ is the thermal stochastic noise in thermal system with zero mean $\langle\xi\rangle=0$. In
the high temperature limit $T\rightarrow\infty$, the noise source is Markovian: $\langle\xi(\mathbf{k},t)\xi(\mathbf{k'},t')\rangle=2\Gamma T(2\pi)^3\delta^3(\mathbf{k}-\mathbf{k'})\delta(t-t')$ \cite{Lisa2004,Gleiser1994}. We can see that the thermal noise term in warm inflation is a kind of Gaussian distributed white noise, and so the leading order inflaton perturbation is pure Gaussian for it's the linear response to the thermal noise. So the field perturbation in warm inflation is Gaussian to good accuracy. Then $\delta N$ formalism can be safely used in calculating the non-Gaussianity generated by warm inflation.

The power spectrum of the curvature perturbation $\zeta$, denoted by $\mathcal{P}_{\zeta}$, is defined as
$\langle\zeta_{\mathbf{k}_1}\zeta_{\mathbf{k}_2}\rangle\equiv(2\pi)^3\delta^3(\mathbf{k}_1+\mathbf{k}_2)
\frac{2\pi^2}{k_1^3}\mathcal{P}_{\zeta}(k_1)$, and $\mathcal{P}_{\zeta}(k)\equiv\frac{k^3}{2\pi^2}P_{\zeta}(k)$. The lowest order non-Gaussianity is three-point function, or its Fourier transform, the bispectrum, which is defined through $\langle\zeta_{\mathbf{k}_1}\zeta_{\mathbf{k}_2}\zeta_{\mathbf{k}_3}\rangle\equiv(2\pi)^3\delta^3
(\mathbf{k}_1+\mathbf{k}_2+\mathbf{k}_3)B_{\zeta}(k_1,k_2,k_3)$. Its normalization is specified by the nonlinear parameter $f_{NL}$ through
\begin{eqnarray}
B_{\zeta}(k_1,k_2,k_3)\equiv - \frac65f_{NL}(k_1,k_2,k_3)\left[P_{\zeta}(k_1)P_{\zeta}(k_2)+cyclic \right].\nonumber \\
&& \label{fnl}
\end{eqnarray}
Observational limits are usually put on this parameter and it can describe the level of non-Gaussianity effectively.

We have mentioned above that $\delta \phi$ in warm inflation is almost Gaussian, so the formula in Eq. (\ref{zeta2}) makes $f_{NL}$ practically independent of wave numbers. The expression of nonlinear parameter is given by \cite{David2005,Boubekeur}
\begin{eqnarray}
-\frac35 f_{NL}=\frac{\sum_{ij}N_{,i}N_{,j}N_{,ij}}{2\left[\sum_i N^2_{,i}\right]^2}+\ln(kL)\frac{\mathcal{P}_{\zeta}}
{2}\frac{\sum_{ijk}N_{,ij}N_{,jk}N_{,ki}}{\left[\sum_i N^2_{,i}\right]^3}.\nonumber \\
&& \label{fNL}
\end{eqnarray}
As discussed in Ref. \cite{David2005,Boubekeur}, the logarithm can be taken to be of order 1 (it involves the size $k^{-1}$ of a typical scale under consideration, relative to the size $L$ of the region within which the stochastic properties are specified), and we know that the power spectrum $\mathcal{P}_{\zeta}$ was fixed by observations with the order ${\cal O}(10^{-9})$, so the leading term of the equation above is the first term. Then if the inflaton fields are Gaussian, $f_{NL}$ is scale independent. If include the possible weak non-Gaussianity of field perturbation $\delta\phi$, we can get additional contribution to the non-Gaussianity of primordial curvature perturbation $\zeta$. The contribution comes from the three-point correlator of the field perturbation and was found to be small in slow roll inflation \cite{Zaballa2005}.

\subsection{\label{sec33}Primordial non-Gaussianity in warm inflation}

There is only one inflaton field in warm inflation, so only one $\delta\phi_i$ is relevant, then Eq. (\ref{zeta2}) becomes
\begin{equation}\label{zeta3}
  \zeta(t,\mathbf{x})=N_{,i}\delta\phi_i+\frac12 N_{,ii}\left(\delta\phi_i\right)^2,
\end{equation}
As we mentioned above, the first term in Eq. (\ref{fNL}) dominates, so Eq. (\ref{fNL}) reduces to
\begin{equation}\label{fNL1}
  -\frac35f_{NL}=\frac12\frac{N_{,ii}}{N^2_{,i}}.
\end{equation}
Since there is only one $\delta\phi_i$, for simplicity, we rewrite $N_{,i}$ as $N_{\phi}$ and $N_{,ii}$ as $N_{\phi\phi}$. In warm inflation, $f_{NL}$ is scale independent for that field perturbation is almost Gaussian.

Through Eq. (\ref{efold}), we can get
\begin{equation}\label{Nphi}
  N_{\phi}=-\frac{1}{M_p^2}\frac{V(1+r)}{V_{\phi}},
\end{equation}

Now we will consider two different warm inflationary cases in slow roll approximation. Observational limits of primordial non-Gaussianity generated by inflation are usually put on the nonlinear parameter. And it's usually estimated on the time of horizon crossing, which is well inside the slow roll inflationary regime.

\begin{itemize}
\item {$\Gamma=\Gamma_0=\textrm{constant}$}
\end{itemize}

For simplicity, a constant dissipation coefficient was often considered in many works. From Eq. (\ref{Nphi}), we can get
\begin{equation}\label{Nphiphi1}
  N_{\phi\phi}=-\frac{1}{M_p^2}\left[(1+r)-\frac{(1+r)\eta}{2\epsilon}-\frac r2\right].
\end{equation}
Finally we can obtain the nonlinear parameter in the case of warm inflation with a constant dissipation coefficient,
\begin{equation}\label{fNL2}
  -\frac35f_{NL}=-\frac{\epsilon}{1+r}+\frac{\eta}{2(1+r)}+\frac{r\epsilon}{(1+r)^2}.
\end{equation}
Considering the slow roll conditions in warm inflation, we can find from the equation above that $|f_{NL}|\sim \mathcal{O}\left(\frac{\epsilon}{1+r}\right)\lesssim1$, which is a first order small quantity in slow roll approximation. For the non-Gaussian level we get by ignoring the non-Gaussianity of field perturbation is not big, we may wonder whether the additional contribution by non-Gaussianity of field perturbation can be important. According to the calculations of \cite{Gupta2002,Gupta2006,Zhang2015}, which concentrate on the three-point correlation functions of the fields, the additional contribution to the nonlinear parameter $f_{NL}$ is a second order small quantity in slow roll approximation. So the non-Gaussian result we get by using $\delta N$ formalism is very effective approximation.

Through Eq. (\ref{fNL2}) we can find that the first-order perturbation contribution to the non-Gaussianity can be quite small and overwhelmed by the second-order
one when horizon crossing and with the expansion of the universe, the contributions form both of them are important and comparable. The level of non-Gaussianity generated by canonical warm inflation is not significant as in canonical cold inflation. The result is well inside the allowed region of Planck \cite{PLANCKNG2015}. Noncanonical warm inflationary models with low sound speed can have significant non-Gaussianity \cite{Zhang2015}.

\begin{itemize}
\item {$\Gamma=\Gamma(\phi)$}
\end{itemize}

The dissipation coefficient is more likely to be a function of the inflaton and even temperature. Temperature dependent case is complicated \cite{MarGil2014}, and we consider temperature independent case here.
We can get
\begin{equation}\label{Nphiphi2}
   N_{\phi\phi}=-\frac{1}{M_p^2}\left[(1+r)-\frac{(1+r)\eta}{2\epsilon}-\frac r2+\frac{r\beta}{2\epsilon}\right]
\end{equation}
from Eq. (\ref{fNL1}). Then we obtain
\begin{eqnarray}
-\frac 35f_{NL} &=&-\frac \epsilon {1+r}+\frac \eta {2(1+r)}+\frac{r\epsilon
}{(1+r)^2}-\frac{r\beta }{2(1+r)^2}.  \nonumber \\
&&  \label{fNL3}
\end{eqnarray}

The result is associate with more slow roll parameters, especially the peculiar parameter $\beta$ in warm inflation. But thanks to the slow roll conditions, the nonlinear parameter is still a first order small quantity in slow roll approximation and is qualitatively the same as the $\Gamma=\Gamma_0$ case.

\section{\label{sec4}Two examples}

\subsection{\label{sec41} Quartic chaotic potential model}
Now we consider a very common model - quartic chaotic potential model, as we all know this model in standard inflation is ruled out by observations of Planck \cite{PLANCKNG2015}. But in warm inflation, the observable predictions from the two-point fluctuations such as power spectrum, spectral index and tensor-to-scalar ratio etc in this model can again be well allowed by Planck observations \cite{Sam2014}, so we take this model as an concrete example to research its three-point fluctuation characters in warm inflation.

The Lagrangian of the inflaton is $\mathcal{L}=\frac12\dot\phi^2-V(\phi)$, with the potential
\begin{equation}\label{quartic}
  V(\phi)=\lambda\phi^4.
\end{equation}
The number of e-folds is given by
\begin{eqnarray}\label{efold1}
  N(\phi)&=&-\frac{1}{M_p^2}\int_{\phi}^{\phi_e}\frac{\phi}4(1+r)d\phi \nonumber \\
  &\simeq &\frac{1+r}{8M_p^2}\left(\phi^2-\phi^2_e\right).
\end{eqnarray}

Now we consider the cases of $\Gamma=\Gamma_0$ and $\Gamma=\Gamma(\phi)$ respectively.

\begin{itemize}
\item {The case of $\Gamma=\Gamma_0$}
\end{itemize}

The slow roll parameters in this case are given by
\begin{equation}\label{SR1}
  \epsilon=\frac{8M_p^2}{\phi^2}<1+r, ~~\eta=\frac{12M_p^2}{\phi^2}<1+r, ~~\beta=0.
\end{equation}
Then from Eq. (\ref{fNL2}), we can get
\begin{equation}\label{fNL4}
  -\frac35f_{NL}=\frac{(6r-2)M_p^2}{(1+r)^2\phi^2}.
\end{equation}
Through Eq. (\ref{efold1}) we can obtain the value of $\phi$ during slow roll inflation
\begin{equation}\label{phistar}
  \phi =2\sqrt{2}M_p\sqrt{\frac{1+N}{1+r}}.
\end{equation}
 As commonly used, the subscript $\ast$ denotes the time when Hubble horizon crossing $k=aH$. In order to solve horizon problem, we need the number of e-folds to be 60 or so. If we set $N=60$, we can see $\phi_{\ast}\simeq2.2M_p$ when $r=100$, $\phi_{\ast}\simeq0.7M_p$ when $r=1000$ and $\phi_e< M_p$ as long as the thermal dissipation effect is not very weak. Quartic potential model is a kind of large field inflationary model which often suffers the problem of overlarge amplitude of inflaton ($\triangle\phi\gg M_p$) in standard inflation. In strong regime of quartic warm inflation model, we can have $\triangle\phi\lesssim M_p$, thus the overlarge amplitude of inflaton suffered in standard inflation is cured in warm inflation.

\begin{center}
\begin{figure}[!ht]
\includegraphics[clip,width=0.48\textwidth]{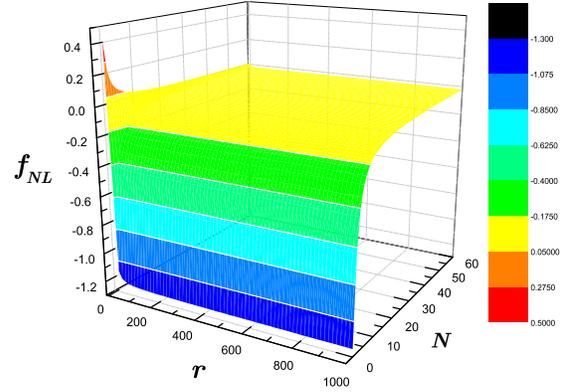}
\caption{(colour online).  The non-linear parameter $f_{NL}$ as a function of dissipation strength of warm inflation $r$ (from weak regime to strong regime) and the number of e-folds $N$.}
\label{Fig1}
\end{figure}
\end{center}

From Eqs. (\ref{fNL4}) and (\ref{phistar}), we can obtain
\begin{equation}\label{fNL5}
  -\frac35f_{NL}=\frac{3r-1}{4(1+r)(1+N)}.
\end{equation}
The equation above shows how the level of non-Gaussianity depends on the dissipation strength and evolve with the expansion of universe. We draw the concrete picture Fig. \ref{Fig1} to show the relation intuitively.

\begin{center}
\begin{figure}[!ht]
\includegraphics[clip,width=0.48\textwidth]{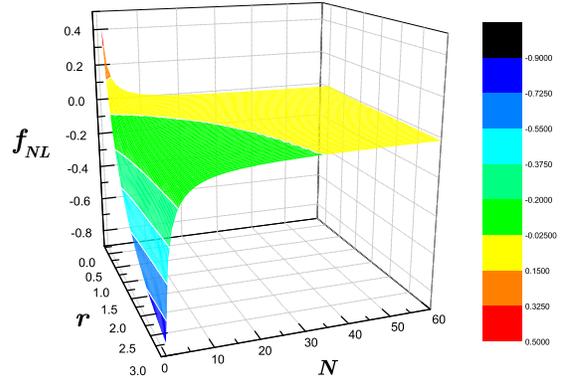}
\caption{(colour online).  The non-linear parameter $f_{NL}$ as a function of dissipation strength of warm inflation $r$ and the number of e-folds $N$ in weak regime of warm inflation.}
\label{Fig2}
\end{figure}
\end{center}

From the Eq. (\ref{fNL5}) and Fig. \ref{Fig1}, we can see $f_{NL}$ changes fast and even changes the sign when the dissipation effect is quite weak, so we plot Fig. \ref{Fig2} to show the details in the region of $0<r<3$. Then the sign of nonlinear parameter remains negative and the magnitude grows slowly.

\begin{center}
\begin{figure}[!ht]
\includegraphics[clip,width=0.48\textwidth]{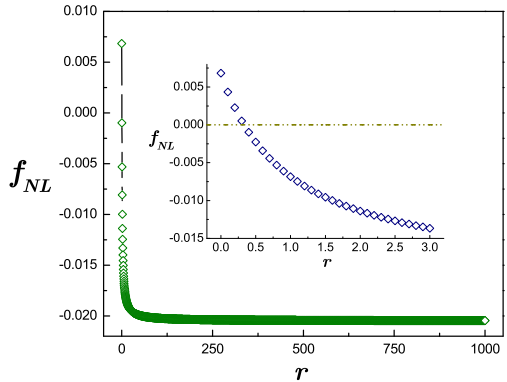}
\caption{ The non-linear parameter $f_{NL}$ as a function of dissipation strength of warm inflation $r$ when cosmological scales exit the hubble horizon ($N=60$ or so).}
\label{Fig3}
\end{figure}
\end{center}

\begin{center}
\begin{figure}[!ht]
\includegraphics[clip,width=0.48\textwidth]{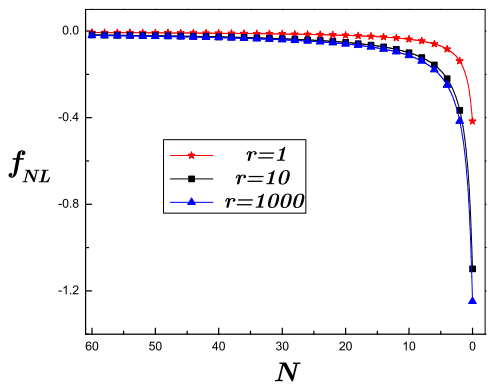}
\caption{(colour online). The non-linear parameter $f_{NL}$ evolves with the expansion of the universe in different regime of warm inflation.}
\label{Fig4}
\end{figure}
\end{center}

In order to see the influence of dissipation strength and the number of e-folds on the level of non-Gaussanity respectively, we draw Figs. \ref{Fig3} and \ref{Fig4} to give a specific and intuitive analysis.

From Fig. \ref{Fig3} we can see that the non-Gaussian effect is very weak when horizon crossing. Nonlinear parameter changes very fast in the range of $0<r<3$ (the clear $r$-dependent curve in the fast changed region is shown by the small picture in Fig. \ref{Fig3}), and changes its sign at $r=\frac1 3$. Then the magnitude of nonlinear increases slowly with the dissipation strength and tends to a limit at $0.02$ or so.

From Fig. \ref{Fig4} we can see that the magnitude of the non-Gaussianity is enhanced with the expansion of the universe (from horizon crossing to the end of inflation) in both strong and weak regime of warm inflation. Non-Gaussian effect is quite weak when cosmological scale crosses the horizon ($N\simeq60$) no matter how large the dissipation strength is, for horizon crossing is well inside the slow roll region. But towards the end of inflation, the level of non-Gaussianity increases fast and strong warm inflation can lead to significant non-Gaussianity somewhat at that time. And with the expansion of the universe, the differences among different dissipative cases become obvious to some degree.

The range of nonlinear parameter in warm inflation given by latest Planck observations is
$f_{NL}^{warm}=-23\pm36$ at $68\%$ CL \cite{PLANCKNG2015}. Our non-Gaussian result in warm quartic inflation was well inside the region allowed by Planck observations as the observable predictions from the two-point fluctuations \cite{Sam2014}. And our result is qualitatively consistent with the negative central value given by Planck.

\begin{itemize}
\item {The case of $\Gamma=\Gamma(\phi)$} \\
\end{itemize}

Now we consider the inflaton dependent case with
\begin{equation}
\Gamma=\Gamma(\phi)\propto C_{\phi}\phi^n,
\end{equation}
where $n$ is inclined to be an integer \cite{Zhangyi2009,ZhangZhu,Campo2010,Lisa2004}.
The slow roll parameters $\epsilon$ and $\eta$ are the same as in Eq. (\ref{SR1}), and
\begin{equation}\label{beta}
  \beta=\frac{4nM_p^2}{\phi^2}.
\end{equation}
The nonlinear parameter in this case is given by
\begin{equation}\label{fNL6}
  -\frac 35 f_{NL}=\frac{M_p^2}{(1+r)^2\phi^2}(6r-2nr-2).
\end{equation}
Using Eq. (\ref{phistar}) we can rewrite the equation above as
\begin{equation}\label{fNL7}
   -\frac 35 f_{NL}=\frac{3r-nr-1}{4(1+r)(1+N)},
\end{equation}
This expression again give out the dependence of non-Gaussianity on dissipation strength and amounts of expansion, which we can see intuitively from Fig. \ref{Fig5}. In order to find the influence of inflaton dependence and make comparison of different cases, we draw the inflaton dependent cases of $n=-2$, $n=-1$, $n=0$, $n=1$, $n=2$ and $n=3$ together in Fig. \ref{Fig5}.

From Fig. \ref{Fig5} we see that when both $r$ and $N$ are small, the differences among them are not obvious and they all have a small positive value of nonlinear parameter. But with the increase of $r$ and $N$, the differences become great. These cases all have a turnaround of the sign of linear parameter except the case of $n=3$, and the magnitude of non-Gaussianity decreases with the integer $n$ increases in the range of $[-2,3]$.

\begin{widetext}
\begin{center}
\begin{figure*}[ht]
\begin{tabular}{ccc}
\includegraphics[width=0.4\textwidth]{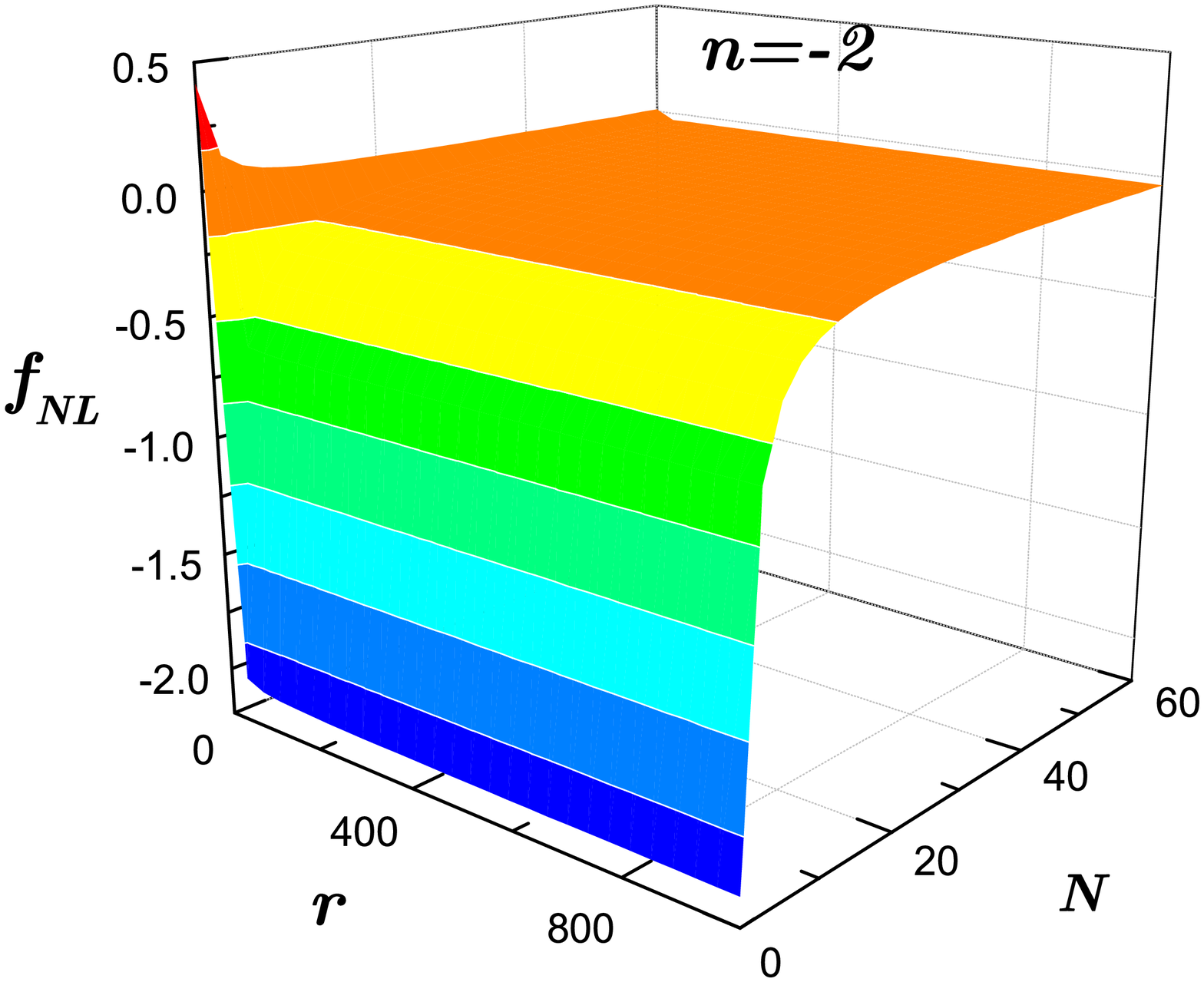}&
\includegraphics[width=0.4\textwidth]{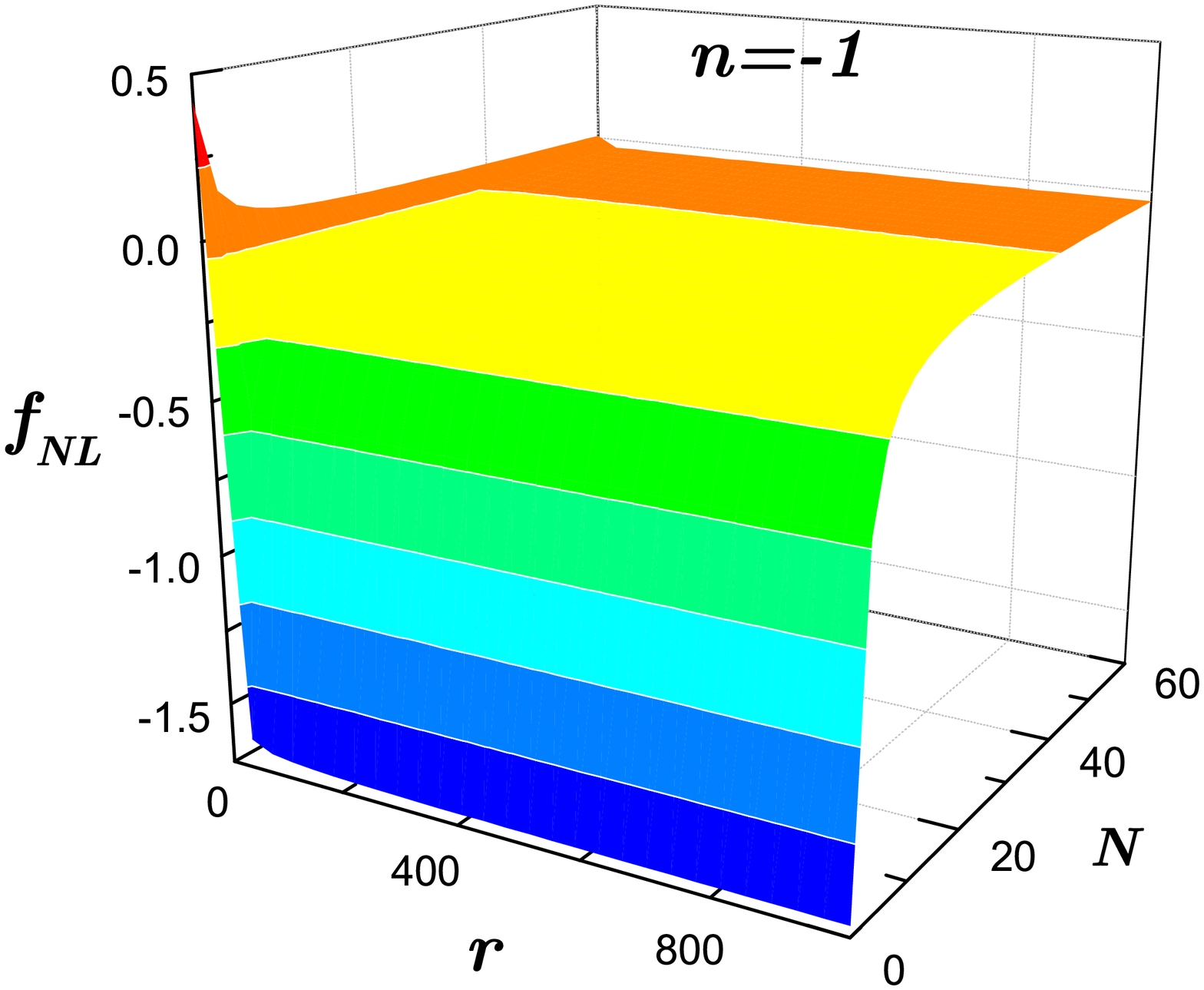}\\
\includegraphics[width=0.4\textwidth]{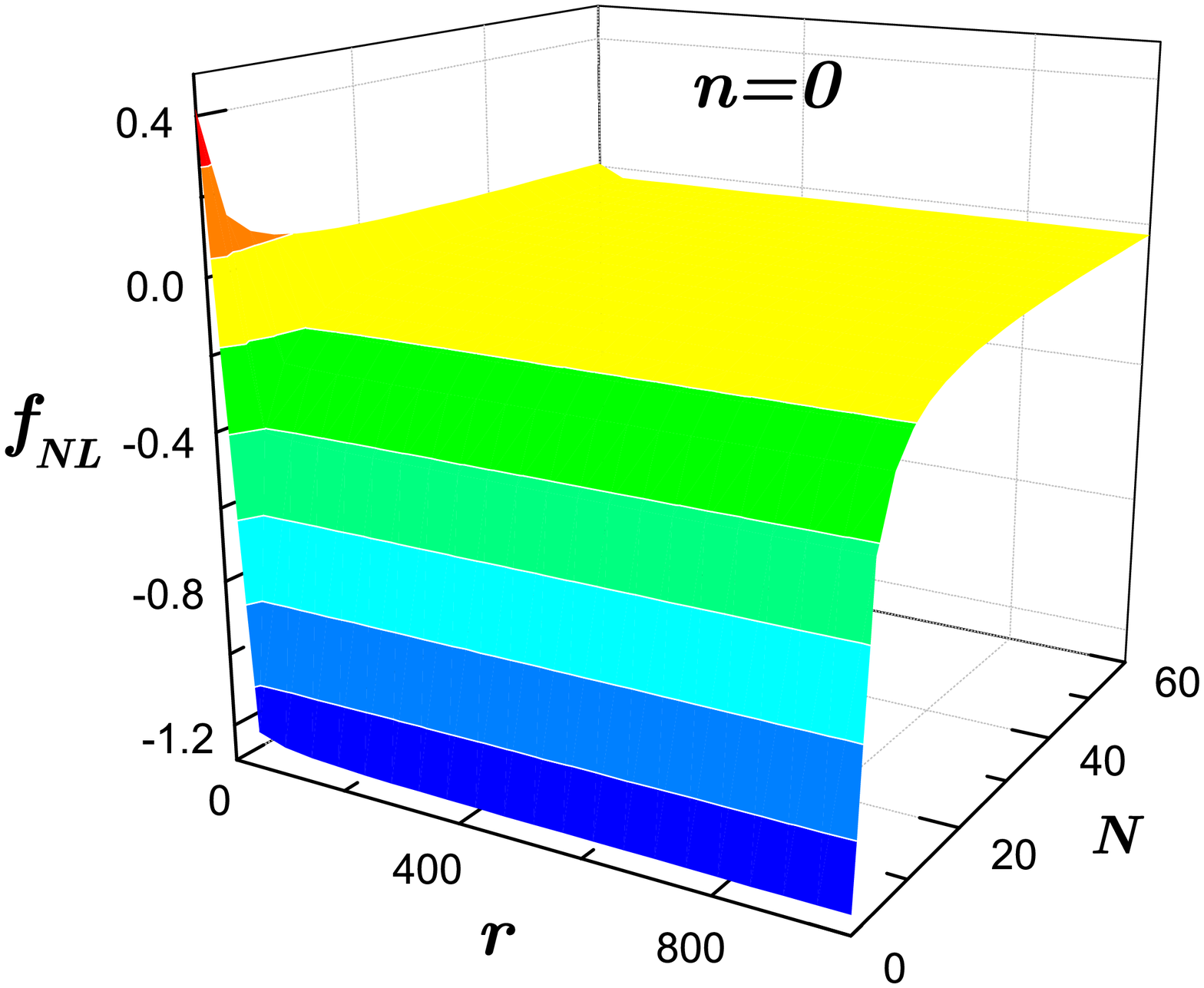}&
\includegraphics[width=0.4\textwidth]{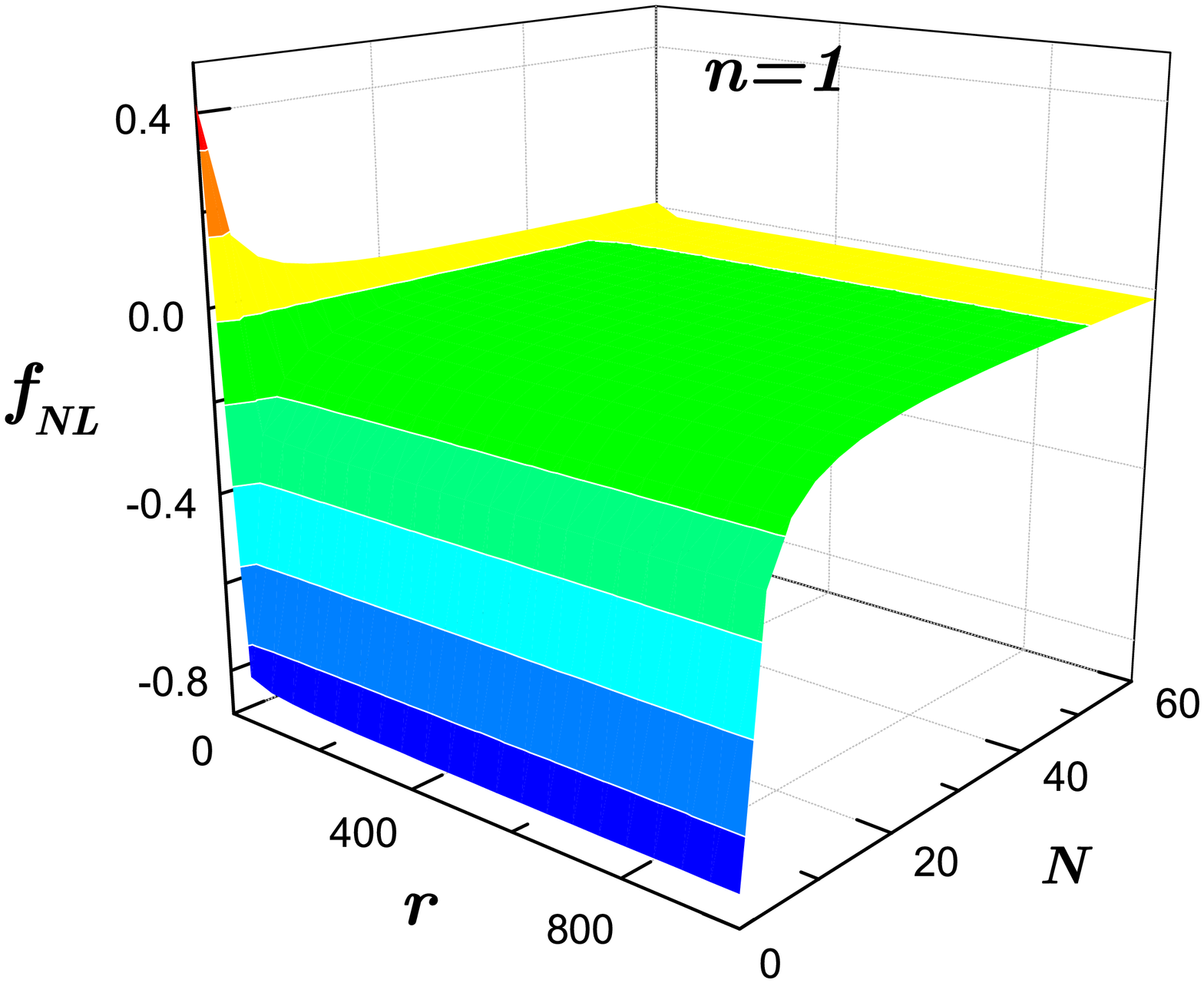}\\
\includegraphics[width=0.4\textwidth]{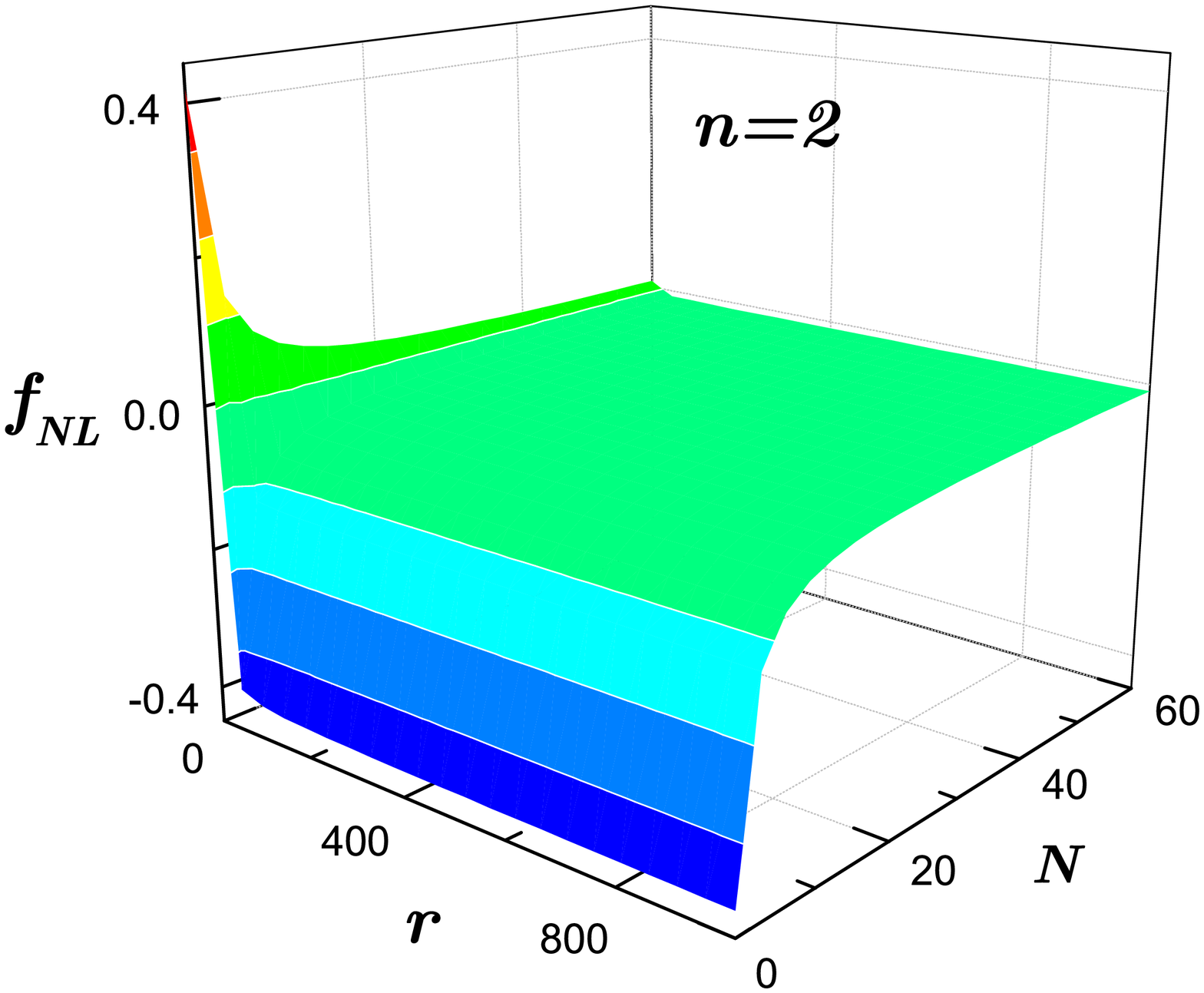}&
\includegraphics[width=0.4\textwidth]{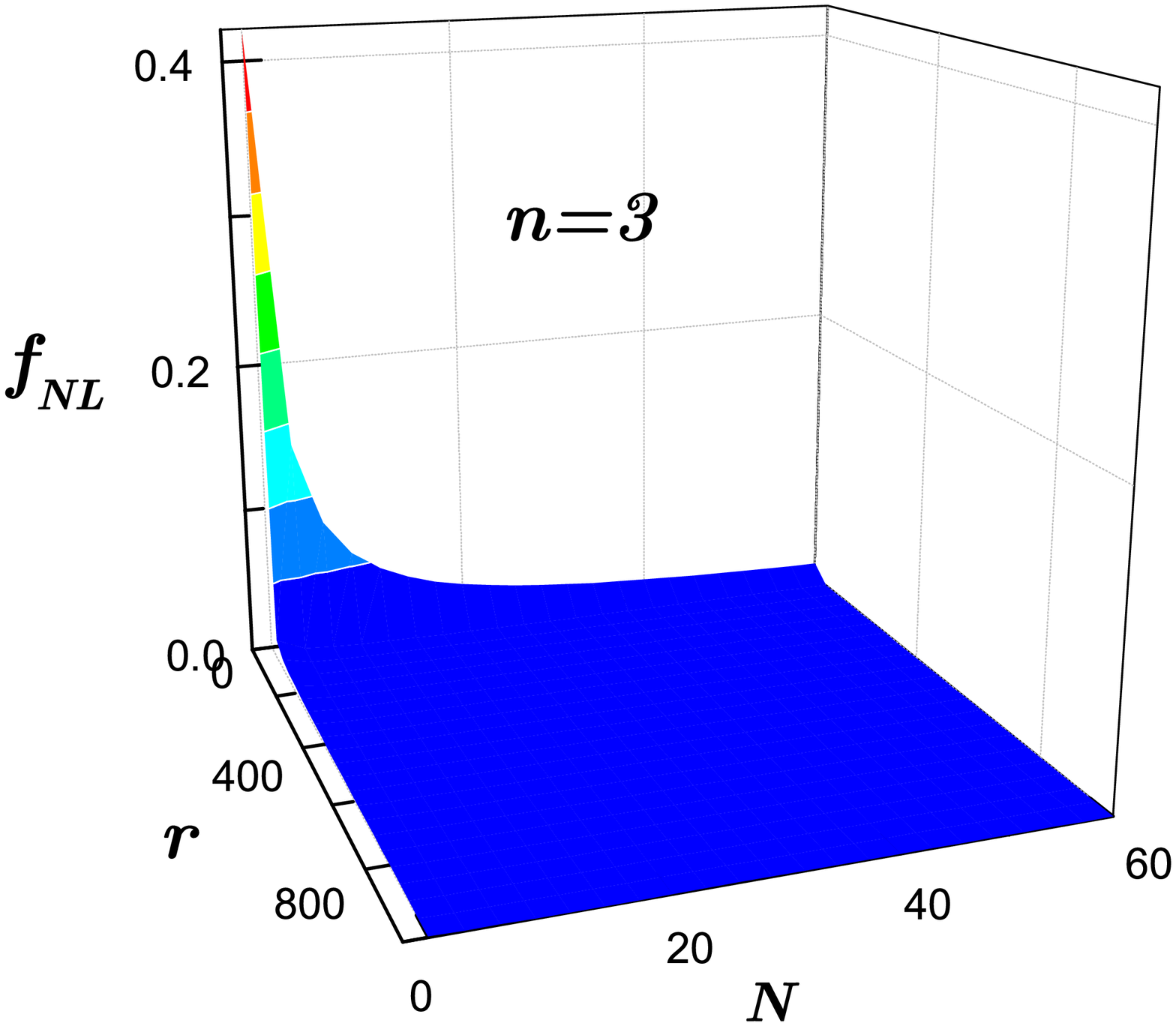}
\end{tabular}\caption{(colour online). The non-linear parameter $f_{NL}$ evolves with the dissipation strength and the amounts of expansion of the universe in different inflaton dependent warm inflationary cases.}
\label{Fig5}
\end{figure*}
\end{center}
\end{widetext}

\subsection{\label{sec42} Hilltop model}
Here we concentrate on the so called hilltop model in warm inflation. The model was proposed in \cite{BoubekeurLyth} and was analysed in both standard inflation \cite{Lin2009,Kohri2007} and warm inflation \cite{Juan2008}. We know $R^2$ model fit the Planck observations best, and hilltop model also fit the Planck observations well \cite{PLANCKI2015}. Hilltop model is a kind of small field model with very flat potential
\begin{equation}\label{hilltopV}
  V=V_0-\frac12m^2\phi^2+\cdots ,
\end{equation}
where $V_0$ is the dominated term ($V\simeq V_0$) and the dots indicates higher order terms in the power series expansion. The inflaton rolls away from an unstable equilibrium.

The slow roll parameters are given by
\begin{equation}\label{SR2}
  \epsilon\cong\frac{M_p^2m^4\phi^2}{2V_0^2}<1+r, \eta\cong\frac{M_p^2m^2}{V_0}<1+r,
\end{equation}
\[
\left\{
\begin{array}{c}
\beta=0, ~~\textrm{when} ~~~\Gamma=\Gamma_0 ;\\
\beta=\frac{-nm^2M_p^2}{V_0}, ~~\textrm{when} ~~~\Gamma=\Gamma(\phi)\propto C_{\phi}\phi^n.
\end{array}
\right.
\]

The number of e-folds is
\begin{equation}\label{efold2}
  N(\phi)\simeq\frac{1}{M_p^2}\int(1+r)\frac{V_0}{m^2\phi}d\phi \simeq\frac{(1+r)V_0}{M_p^2m^2}\ln\left(\frac{\phi_e}{\phi}\right).
\end{equation}
When inflation ends at $\epsilon=1+r$, we can get $\phi_e=\sqrt{2(1+r)}V_0/(M_p^2 m^2)$, which is bigger than that in standard inflation, so inflation can last longer in warm inflation than in standard inflation since this is a kind of small field model.

In the $\Gamma=\Gamma_0$ case, from Eqs. (\ref{fNL2}) and (\ref{SR2}), we can get
\begin{equation}\label{fNL8}
  -\frac35f_{NL}=\frac{M_p^2m^2}{2V_0(1+r)}\left[1-\frac{1}{1+r}\frac{m^2\phi^2}{V_0}\right],
\end{equation}
As the slow roll condition $\eta\ll1+r$ indicates, the first factor in front of the bracket is much less than 1, especially when horizon crossing. The second term in the bracket of Eq. (\ref{fNL8}) is also tiny, so we can obtain $|f_{NL}|\sim\mathcal{O}\left[\frac{M_p^2m^2}{V_0(1+r)}\right]\ll1$ in warm hilltop inflation with $\Gamma=\Gamma_0$. The tiny non-Gaussianity in this model is due to the very flat small field potential and the added big thermal damped term in the evolution equation of inflaton.

In the $\Gamma=\Gamma(\phi)\propto C_{\phi}\phi^n$ case, we have
\begin{equation}\label{fNL9}
  -\frac35f_{NL}\doteq\frac{M_p^2m^2}{2V_0(1+r)}\left[1+\frac{nr}{1+r}\right].
\end{equation}
For the same reason, the first factor in the equation above is much less than 1, and the second factor is of order unity, so the nonlinear parameter is still quite small. If $|n|$ is much larger than 1, the magnitude of non-Gaussianity can be enhanced to a great extent compared to the case of $\Gamma=\Gamma_0$, but $|n|$ cannot be too large. On the whole, non-Gaussianity in this model is still insignificant for the same reason as in $\Gamma=\Gamma_0$ case.

\section{\label{sec5}conclusions}

In this paper, A $\delta N$ formalism is used to investigate the non-Gaussianity generated in warm inflation. The formalism is convenient to use and often used in calculating the non-Gaussianities in multi-field inflation theories. We give a brief introduction of the warm inflation, and then introduce the $\delta N$ formalism in detail. The inflaton perturbation in warm inflation is almost Gaussian, so the calculation using $\delta N$ formalism is valid to sufficient accuracy. Warm inflation is dominated by one inflaton field, so we use the $\delta N$ formalism that reduces to single field case. Non-Gaussianity is conveniently described by the nonlinear parameter $f_{NL}$ and observational limits of primordial non-Gaussianity generated by inflation are usually put on this parameter. We find that the nonlinear parameter in single-field warm inflation is scale-independent. Using $\delta N$ formalism, we obtain the expression of nonlinear parameter in two cases of warm inflation: $\Gamma=\Gamma_0$ and $\Gamma=\Gamma(\phi)$ cases. In both cases, the nonlinear parameter can be expressed as a linear combination of the slow roll parameters, so it's a first order small quantity in
slow roll approximation with the order $|f_{NL}|\sim \mathcal{O}\left(\frac{\epsilon}{1+r}\right)$. That indicates the non-Gaussianity generated by warm inflation is insignificant especially when horizon crossing as in canonical single-field standard inflation.

In order to find how the nonlinear parameter evolves and which parameters it depends, we study two examples to give out concrete pictures. The first example is the quartic chaotic potential model and the second one is the hilltop model. We get specific results in the $\Gamma=\Gamma_0$ and $\Gamma=\Gamma(\phi)$ cases in both examples. The nonlinear parameter depends on the dissipation strength of the warm inflation and the amounts of expansion of the universe. We study the quartic warm inflation with $\Gamma=\Gamma_0$ in details. Through this example, we find that the nonlinear parameter changes its sign when $r$ is a quite small and then it's always negative with the increase of $r$. The magnitude of the nonlinear parameter increases with $r$ very fast in a weak regime of the warm inflation, and then it increases slowly with $r$ and tends to a limit. The non-Gaussianity is tiny when the horizon crossing and can be enhanced with expansion of the universe. But the level of the non-Gaussianity is still not very significant at the end of the inflation in canonical quartic warm inflation. The non-Gaussian results we get are well inside the allowed region of Planck observations. Then we discuss the warm hilltop model and find that this model predicts a quite weak non-Gaussianity thanks to the very flat potential and added thermal damped term in both $\Gamma=\Gamma_0$ and $\Gamma=\Gamma(\phi)$ cases. But the non-Gaussianty generated in inflaton dependent case is enhanced to some degree compared to $\Gamma=\Gamma_0$ case.

We find that the non-Gaussianity generated by the warm inflation is not very significant, for it's a kind of canonical single-field inflation theory. The extension of the warm inflation - noncanonical warm inflation was proposed in \cite{Zhang2014} and its non-Gaussian problem was partly discussed in \cite{Zhang2015}. We'll use the $\delta N$ formalism to investigate the issue of the non-Gaussianity in noncanonical warm inflation in the future to see whether it can predict a high level non-Gaussianity.

%----------------------------------------------------------------------------------------------------
\acknowledgments Xiao-Min Zhang, Hong-Yang Ma and Peng-Cheng Chu were supported by the National Natural Science Foundation of China (Grants No. 61572270, 61173056, 11547035, 11304147 and 11505100) and the Shandong Provincial Natural Science Foundation, China (ZR2015AQ007). Xiao-Min Zhang and Jian-Yang Zhu were supported by the National Natural Science Foundation of China (Grants No. 11175019 and 11235003).
%---------------------------------------------------------------------------------------------------------

\end{document}